\documentclass[prd,amsmath,amssymb,showpacs,superscriptaddress,nofootinbib,twocolumn]{revtex4-1}
\usepackage{graphicx}
\usepackage{epsfig,graphics,subfigure,psfrag,amsmath,amssymb}
\usepackage{lineno}
\usepackage{dcolumn}
\usepackage{bm}
\usepackage{overpic}
\usepackage{xspace}

\usepackage{color}
\usepackage{multirow}
\usepackage{colortbl}
\usepackage{epstopdf}
\usepackage[colorlinks,linkcolor=blue,anchorcolor=blue,citecolor=blue]{hyperref}

\newcommand{\pip}{\pi^{+}}
\newcommand{\pim}{\pi^{-}}

\newcommand{\BESIII}{BES\uppercase\expandafter{\romannumeral3}\xspace}

\newcommand\Tstrut{\rule{0pt}{2.4ex}}

\begin{document}
\title{\boldmath Study of two-photon decays of pseudoscalar mesons\\via $J/\psi$ radiative decays}

\author{M.~Ablikim$^{1}$, M.~N.~Achasov$^{9,d}$, S. ~Ahmed$^{14}$, M.~Albrecht$^{4}$, A.~Amoroso$^{53A,53C}$, F.~F.~An$^{1}$, Q.~An$^{50,40}$, J.~Z.~Bai$^{1}$, Y.~Bai$^{39}$, O.~Bakina$^{24}$
% ...
\\
      \vspace{0.2cm}
      (BESIII Collaboration)\\
      \vspace{0.2cm} {\it
$^{1}$ Institute of High Energy Physics, Beijing 100049, People's Republic of China\\
% ...
$^{4}$ Bochum Ruhr-University, D-44780 Bochum, Germany\\
% ...
$^{9}$ G.I. Budker Institute of Nuclear Physics SB RAS (BINP), Novosibirsk 630090, Russia\\
% ...
$^{14}$ Helmholtz Institute Mainz, Johann-Joachim-Becher-Weg 45, D-55099 Mainz, Germany\\
% ...
$^{24}$ Joint Institute for Nuclear Research, 141980 Dubna, Moscow region, Russia\\
% ...
$^{39}$ Southeast University, Nanjing 211100, People's Republic of China\\
$^{40}$ State Key Laboratory of Particle Detection and Electronics, Beijing 100049, Hefei 230026, People's Republic of China\\
% ...
$^{50}$ University of Science and Technology of China, Hefei 230026, People's Republic of China\\
% ...
$^{53}$ (A)University of Turin, I-10125, Turin, Italy; (B)University of Eastern Piedmont, I-15121, Alessandria, Italy; (C)INFN, I-10125, Turin, Italy\\
% ...
$^{d}$ Also at the Novosibirsk State University, Novosibirsk, 630090, Russia\\
% ...
}
}
\noaffiliation{}

\date{\today}

\begin{abstract}
  Using a sample of $4.48\times10^{8}$~$\psi(3686)$ events collected
  with the BESIII detector at the BEPCII collider, we study the
  two-photon decays of the pseudoscalar mesons $\pi^0$, $\eta$,
  $\eta^\prime$, $\eta(1405)$, $\eta(1475)$, $\eta(1760)$, and
  $X(1835)$ in $J/\psi$ radiative decays using
  $\psi(3686)\to\pi^{+}\pi^{-}J/\psi$ events. The $\pi^0$, $\eta$ and
  $\eta^\prime$ mesons are clearly observed in the two-photon mass
  spectra, and the branching fractions are determined to be
  $B(J/\psi\to\gamma\pi^{0}\to3\gamma)=(3.57\pm0.12\pm0.16)\times10^{-5}$,
  $B(J/\psi\to\gamma\eta\to3\gamma)=(4.42\pm0.04\pm0.18)\times10^{-4}$,
  and
  $B(J/\psi\to\gamma\eta'\to3\gamma)=(1.26\pm0.02\pm0.05)\times10^{-4}$,
  where the first errors are statistical and the second systematic.
  No clear signal for $\eta(1405)$, $\eta(1475)$, $\eta(1760)$
  or $X(1835)$ is observed in the two-photon mass spectra, and upper limits at the
  $90\%$ confidence level on the product branching fractions
  are obtained.
\end{abstract}

\pacs{13.66.Bc, 14.40.Be}
\maketitle

\section{INTRODUCTION}

Within the framework of quantum chromodynamics (QCD), the two-photon
decay width of a meson plays a crucial role in understanding the nature of the meson
and helps to distinguish glueballs from conventional
mesons since glueballs are believed to have a relatively small two-photon decay width~\cite{twophotontheory}.
 Therefore, experimental studies of the
two-photon decays of mesons are very important to help interpret the
meson spectrum.

The $\eta(1405)/\eta(1475)$ pseudoscalar meson was once regarded as a
glueball candidate since it was copiously produced in $J/\psi$
radiative decays~\cite{song18} and was not observed
in
two-photon collisions~\cite{song19}. However, the measured mass
is much lower than the prediction of Lattice QCD for a pseudoscalar glueball,
which lies above 2GeV/$c^{2}$~\cite{LQCD1,LQCD2}.
Later,  the experiments found two different pseudoscalar states,
$\eta(1405)$ and $\eta(1475)$, with the former mainly decaying to $a_{0}(980)\pi$ and $K\bar K\pi$
and the latter mainly to $K^{*}(892)\bar K$~\cite{PDG}.
At present, the one state assumption and the nature of $\eta(1405)$/$\eta(1475)$ are still controversial.
Another pseudoscalar meson, the $\eta(1760)$, has been proposed as a
mixture of a glueball with a conventional $q \bar q$
state~\cite{1760N19}, rather than a pure $q \bar q$ meson or a
glueball, and this hypothesis is supported by the large production
rate of the $\eta(1760)$ in $J/\psi\to\gamma\omega\omega$
decays~\cite{1760N18,1760N18-2}.  The nature of the $X(1835)$ is still
an open question although a number of theoretical interpretations have
been proposed, including an $N \bar{N}$ bound state~\cite{16},
baryonium with sizable gluon content~\cite{17}, a pseudoscalar
glueball~\cite{18}, a radial excitation of the $\eta'$~\cite{19}, and
an $\eta_{c}$-glueball mixture~\cite{20}. None of these
interpretations have been completely ruled out or confirmed.

Pseudoscalar mesons are copiously produced in $J/\psi$ radiative
decays. The two-photon decay widths of $\pi^0$, $\eta$ and
$\eta^\prime$ mesons have been measured~\cite{PDG}, and previous
values were used to determine the branching fractions of
$J/\psi\rightarrow\gamma(\pi^0,~\eta,$
$\eta^\prime)$~\cite{BESII,CLEO}. Those of
$J/\psi\rightarrow\gamma(\eta,$ $\eta^\prime)$ were then used to
calculate the pseudoscalar mixing angle~\cite{BESII}. However, the
two-photon decays of $\eta(1405)$, $\eta(1475)$, $\eta(1760)$ and
$X(1835)$ have not been investigated yet.

At present, the sample of $4.48\times10^{8} ~\psi(3686)$
events~\cite{NumberOfpsip0912} ($1.06\times10^{8}$
events in 2009 and $3.41\times10^{8}$ in 2012)
collected by the BESIII detector offers the opportunity to study the
two-photon decays of pseudoscalar mesons in $J/\psi$ radiative decay
in $\psi(3686)\to\pi^{+}\pi^{-}J/\psi$ events.  While the number of
$J/\psi$ events from the BESIII $\psi(3686) \to \pi^{+}\pi^{-} J/\psi$ data
samples is much smaller than that of the direct BESIII $e^+e^-
\to J/\psi$ samples, the direct $J/\psi$ samples have a large background from the
$e^{+}e^{-}\to\gamma\gamma$ process.  Thus, better sensitivity on the
two-photon decay widths of pseudoscalar mesons is possible using the
$\psi(3686)$ data samples collected at BESIII.  In this paper, the
branching fractions of
$J/\psi\to\gamma(\pi^{0},~\eta,~\eta')\to3\gamma$ are
measured. Additionally, we also search for the two-photon decays of
the pseudoscalar mesons, $\eta(1405)$, $\eta(1475)$, $\eta(1760)$ and
$X(1835)$.

\section{DETECTOR AND MONTE CARLO SIMULATION}

BEPCII is a double-ring $e^{+}e^{-}$ collider running at
center-of-mass energies from 2.0 to 4.6 GeV.  The
BES\uppercase\expandafter{\romannumeral3}~\cite{Ablikim2010345}
detector at BEPCII, with a geometrical acceptance of 93\% of 4$\pi$
solid angle, operates in a 1.0~T magnetic field provided by a superconducting solenoid magnet. The
detector is composed of a helium-based drift chamber (MDC), a
plastic-scintillator time-of-flight (TOF) system, a CsI(Tl)
electromagnetic calorimeter (EMC) and a resistive plate chamber
(RPC)-based muon chamber (MUC) in the iron flux return yoke of the
magnet.  The spatial resolution of the MDC is
better than 130 $\mu$m, the charged-particle momentum resolution is
0.5\% at 1.0 GeV/$c$, and the specific energy loss ($dE/dx$) resolution
is better than 6\% for electrons from Bhabha events.  The time
resolution of the TOF is 80~ps in the barrel and 110~ps in the
endcaps. The energy resolution of the EMC at 1.0~GeV/$c$ is 2.5\% (5\%)
in the barrel (endcaps), and the position resolution is better than
6~mm (9~mm) in the barrel (endcaps).  The position resolution in the MUC
is better than 2~cm.

Monte Carlo (MC) simulations are used to estimate backgrounds and
determine the detection efficiencies. The
GEANT4-based~\cite{ref:geant4} simulation software
BOOST~\cite{ref:boost} includes the geometric and material description
of the BESIII detector, detector response, and digitization models, as
well as the tracking of the detector running conditions and
performance.  Production of the charmonium state $\psi(3686)$ is
simulated with
KKMC~\cite{ref:kkmc,ref:kkmc2}, while the decays are generated with
EVTGEN~\cite{ref:evtgenPing,ref:evtgen} for known decay modes with branching
fractions taken from the Particle Data Group (PDG)~\cite{PDG} and by
LUNDCHARM~\cite{ref:lundcharm} for the remaining unknown decays.  We
use a sample of $5.06 \times 10^{8}$ simulated $\psi(3686)$ events,
in which the $\psi(3686)$ decays generically (`inclusive MC sample'),
to study the backgrounds.
The analysis is
performed in the framework of the BESIII offline software system
(BOSS)~\cite{ref:boss} which incorporates the detector calibration,
event reconstruction, and data storage.

\section{Data analysis}
In this paper, the two-photon decays of the pseudoscalar mesons
are investigated with $J/\psi$ radiative
decays. Hence the candidate events for the reconstruction of
$\psi(3686)\to\pi^{+}\pi^{-}J/\psi$, $J/\psi\to 3 \gamma$ are required
to have two oppositely charged tracks and at least three photon candidates.
Each charged track, reconstructed using hits in the MDC, is required
to be in the polar angle range $|\cos\theta| < 0.93$ and pass the
interaction point within $\pm10$~cm along the beam direction, and
within $\pm1$~cm in the plane perpendicular to the beam. Both
charged tracks are assumed to be pion candidates.

Photon candidates are reconstructed from clusters of energy deposited
in the EMC, and the deposited energy of each is required to be larger
than 25~MeV in the barrel region ($|\cos\theta|<0.80$) or 50 MeV in
the endcap region ($0.86<|\cos\theta|<0.92$). The opening angle between a shower
and the nearest charged track must be greater than $15^{\circ}$, and
timing requirements in the EMC are used to suppress electronic noise
and energy deposits unrelated with the collision event.  Events that satisfy the
above requirements are retained for further analysis.

\begin{figure}[htbp]
  \centering
  \vskip -0.2cm
  \hskip -0.4cm \mbox{
  \begin{overpic}[width=0.4\textwidth]{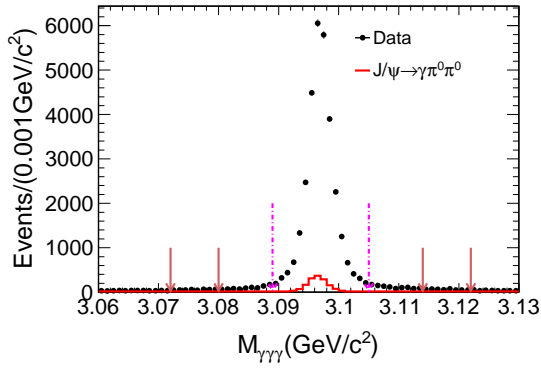}
  \put(80,60){{\bf  }}
  \end{overpic}
  }

  \caption{\label{3gamma} (color online) Three-photon invariant mass spectrum $M_{\gamma\gamma\gamma}$  for
    data (dots with error bars) and MC simulation of
    the background contribution from $J/\psi\to\gamma\pi^{0}\pi^{0}$ (red solid histogram). The pink dot-dashed arrows indicate the signal region for selection of
    $J/\psi$ events, and the brown solid arrows show the
    sideband regions.}
  \end{figure}

A four-constraint (4C) kinematic fit imposing energy and momentum
conservation is performed under the hypothesis of
$\pip\pim\gamma\gamma\gamma$. If the number of photon candidates in an
event is larger than
three, the combination with the smallest $\chi_{4C}^{2}$ from the
kinematic fit is selected,
and $\chi_{4C}^{2}$ is further required to be less than 50.  The
distribution of the $\gamma\gamma\gamma$ invariant mass,
$M_{\gamma\gamma\gamma}$, of selected candidate events is shown in
Fig.~\ref{3gamma}, where a very clean $J/\psi$ peak is seen with very
low background.
A mass window requirement $|M_{\gamma\gamma\gamma} -
m_{J/\psi}| < 0.08$~GeV/$c^{2}$ is applied to select the $J/\psi$ signal, where $m_{J/\psi}$ is
the nominal mass of the $J/\psi$ meson~\cite{PDG}.

\begin{figure}[htbp]
  \centering
  \mbox{
  \begin{overpic}[width=0.4\textwidth]{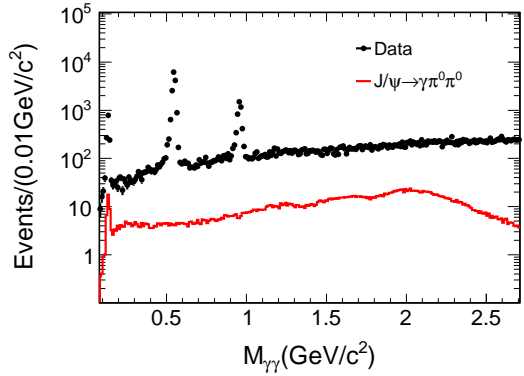}
  \put(80,60){{\bf  }}
  \end{overpic}
  }
  \caption{\label{2gamma}(color online) Two-photon invariant
    mass spectrum for data (dots with error bars) and MC simulation of
    $J/\psi\to\gamma\pi^{0}\pi^{0}$ (red solid histogram).}
\end{figure}

  After the above requirements, the distribution of the two-photon
  invariant mass $M_{\gamma\gamma}$ is shown in
  Fig.~\ref{2gamma}, where the photon momenta from the 4C kinematic
  fit are used to calculate $M_{\gamma\gamma}$ and there are three
  entries per event.

The backgrounds without the $J/\psi$ intermediate state (non-$J/\psi$
background) can be estimated from the events within the $J/\psi$
sideband regions, defined as 3.072~GeV/$c^{2} < M_{\gamma\gamma\gamma}
< 3.080$~GeV/$c^{2}$ and 3.114~GeV/$c^{2} < M_{\gamma\gamma\gamma} <
 3.122$~GeV/$c^{2}$, which are indicated in Fig.~\ref{3gamma}.  The
backgrounds from $\psi(3686)\to\pi^{+}\pi^{-}J/\psi$ with $J/\psi$
decaying to neutral particle final states ($J/\psi$ background) are
investigated with the inclusive MC sample of $5.06\times10^{8}$
$\psi(3686)$ events.  One prominent background is
$\psi(3686)\to\pi^{+}\pi^{-}J/\psi$, with
$J/\psi\to\gamma\pi^0\pi^0$, which produces a peak around the
$\pi^{0}$ mass region in the $M_{\gamma\gamma}$ distribution.
To estimate its contribution, a dedicated MC sample of
$\psi(3686)\to\pi^{+}\pi^{-}J/\psi$, $J/\psi\to\gamma\pi^{0}\pi^{0}$
is produced incorporating the amplitude analysis result of
$J/\psi\to\gamma\pi^{0}\pi^{0}$~\cite{PWAgammapi0pi0}.  With the same
selection criteria and taking into account the number of $\psi(3686)$
events as well as the branching fractions of
$\psi(3686)\to\pi^{+}\pi^{-}J/\psi$~\cite{PDG} and
$J/\psi\to\gamma\pi^{0}\pi^{0}$~\cite{PWAgammapi0pi0}, the
corresponding distribution of $M_{\gamma\gamma}$ is shown as the solid
histogram in Fig.~\ref{2gamma}.  The number of peaking background
events in the $\pi^{0}$ signal region is expected to be $32\pm2$,
which is estimated by a fit to the ${\gamma\gamma}$ invariant mass
spectrum of the above MC sample, where the $\pi^{0}$ signal is modeled
with the sum of a Crystal Ball (CB)~\cite{CBfunction} function and a Gaussian function,
and the other $J/\psi$ non-peaking background is described with a
second order Chebychev polynomial function.

\begin{figure}[htbp]
  \centering
  \mbox{
  \begin{overpic}[width=0.4\textwidth]{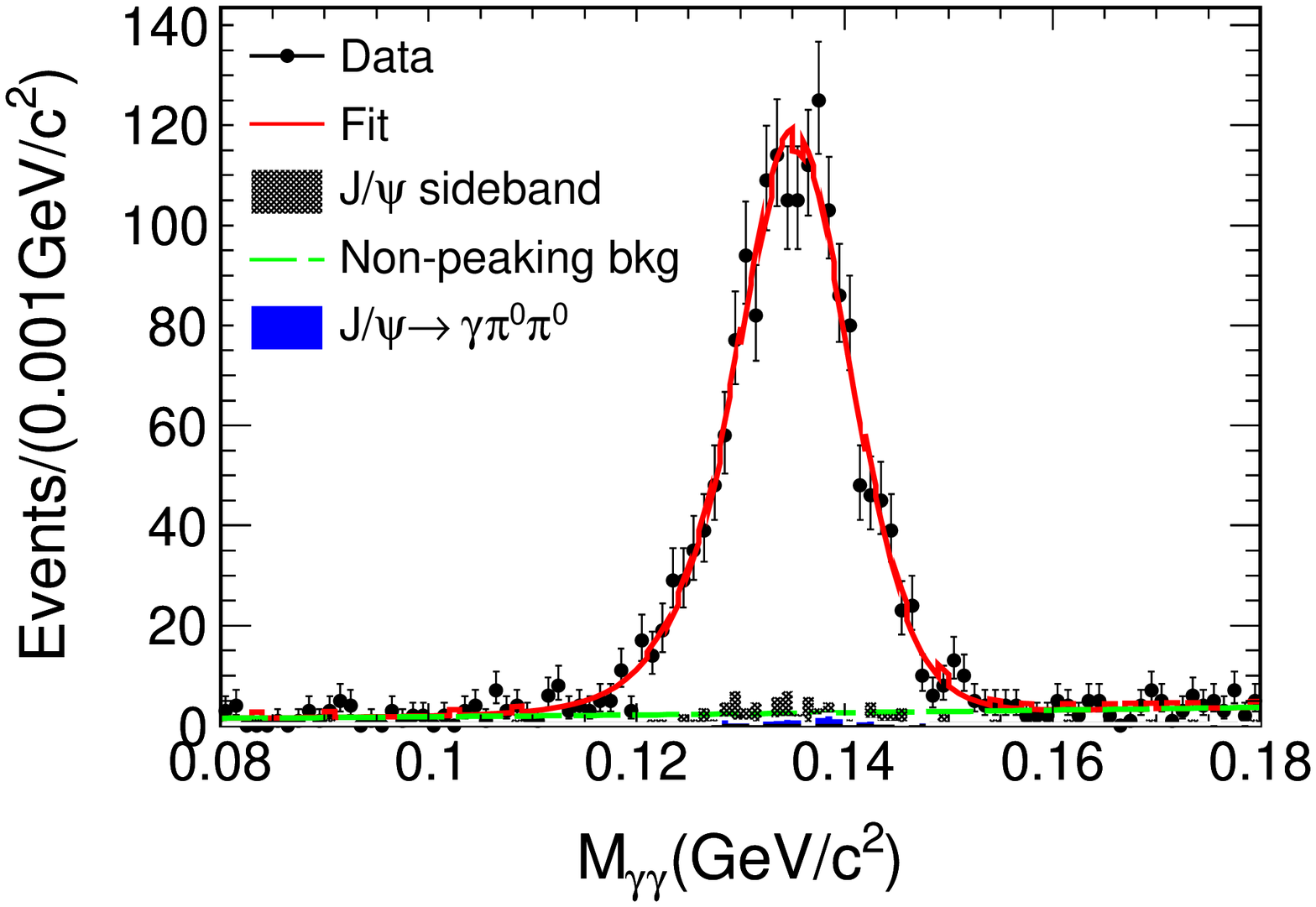}
  \put(80,60){{\bf(a)  }}
  \end{overpic}}

  \mbox{
  \begin{overpic}[width=0.4\textwidth]{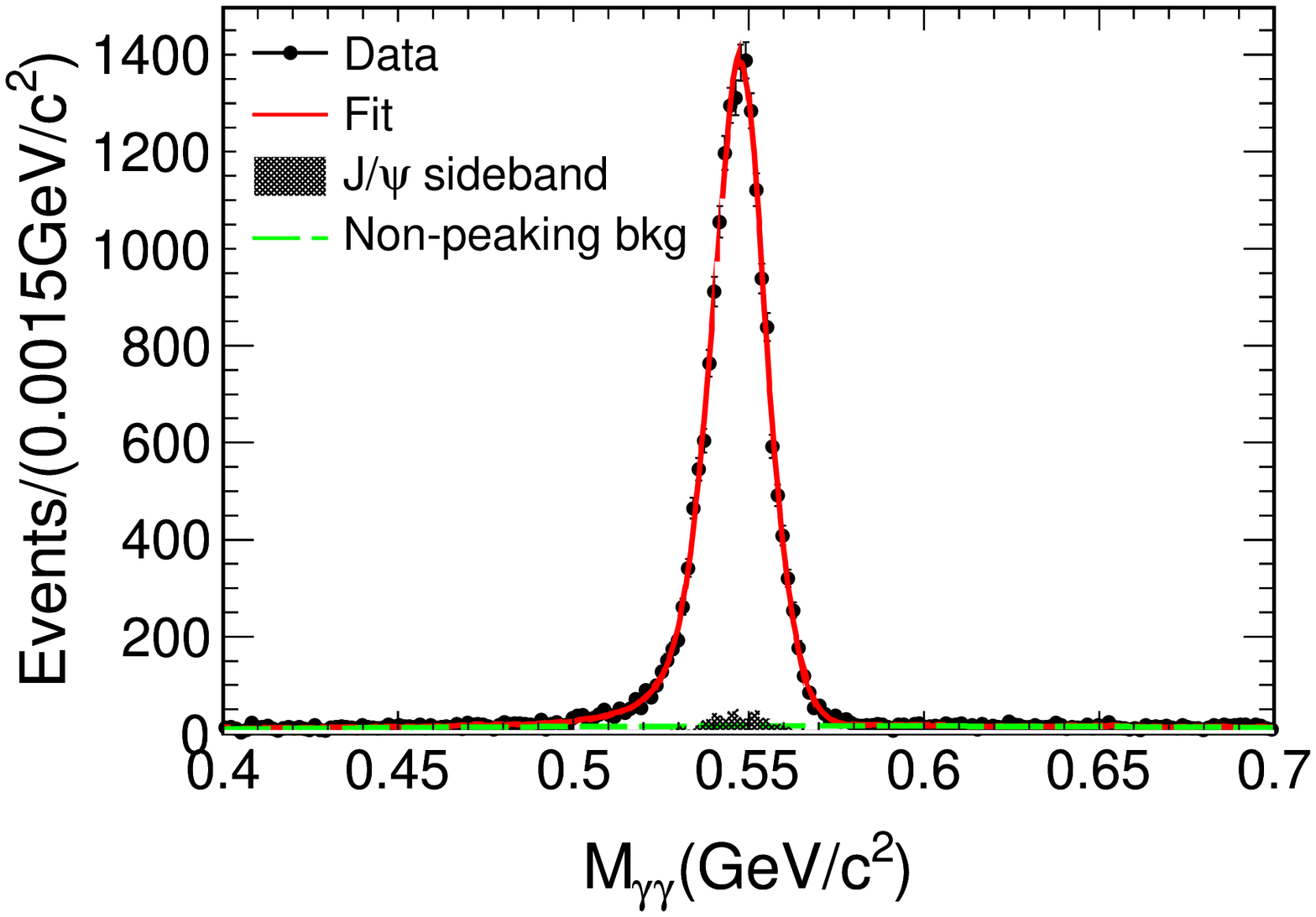}
  \put(80,60){{\bf(b) }}
  \end{overpic}}

  \mbox{
  \begin{overpic}[width=0.4\textwidth]{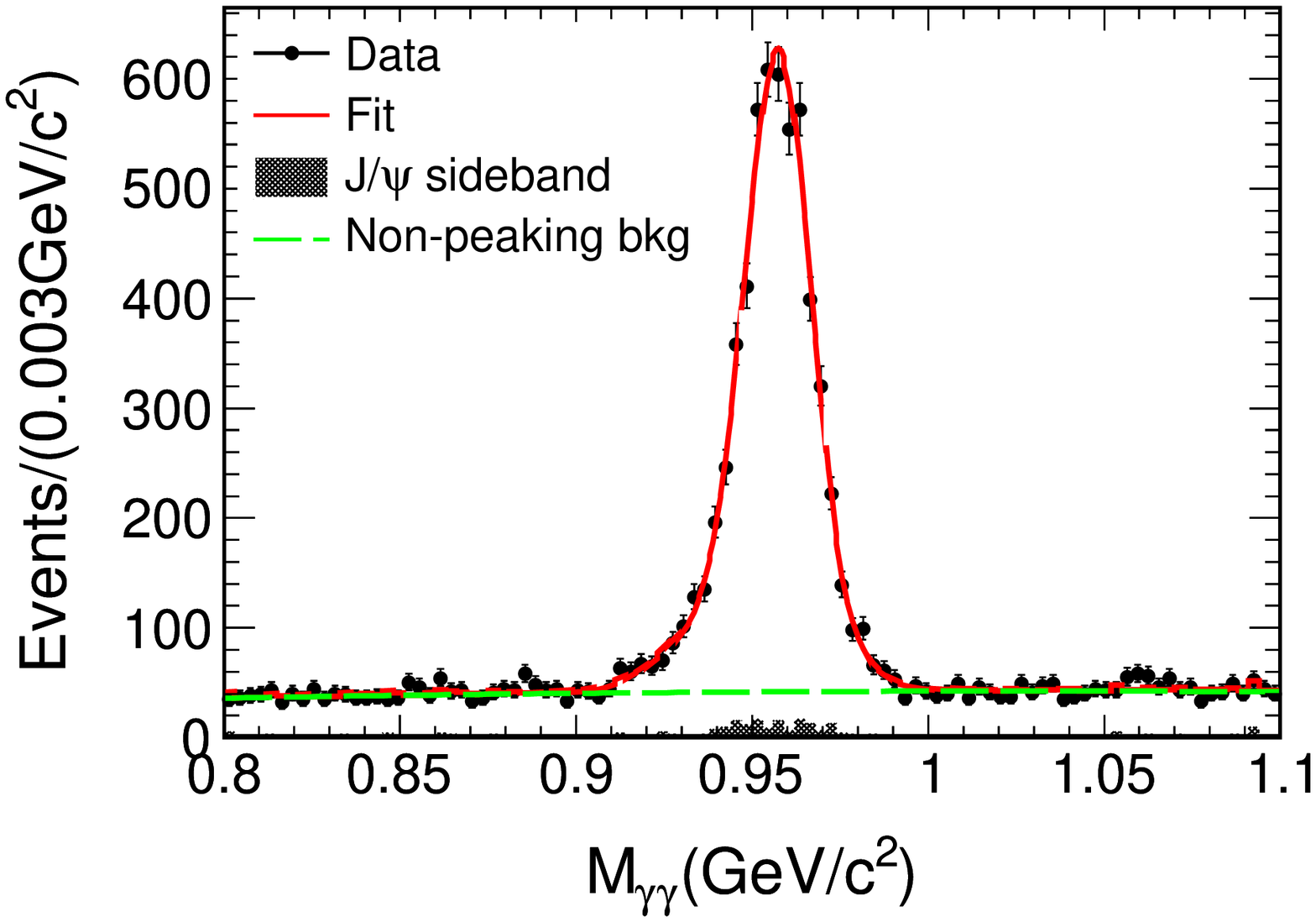}
  \put(80,60){{\bf(c) }}
  \end{overpic}}
  \caption{\label{fitMCpi0}(color online) Fits to the $\gamma\gamma$ mass
    distribution for (a) $J/\psi\to\gamma\pi^{0}\to3\gamma$, (b)
    $J/\psi\to\gamma\eta\to3\gamma$ and (c)
    $J/\psi\to\gamma\eta'\to3\gamma$.  The dots with error bars are
    data; the red solid curve is the result of the fit; the black hatched
    histogram shows the $J/\psi$ sideband background; the
    long-dashed curve represents the other non-peaking backgrounds;
    the blue solid histogram in (a) represents the contribution from the
    background of $J/\psi\to\gamma\pi^{0}\pi^{0}$.}
\end{figure}

 \begin{figure*}[htbp]
  \centering
  \mbox{
  \begin{overpic}[width=0.4\textwidth]{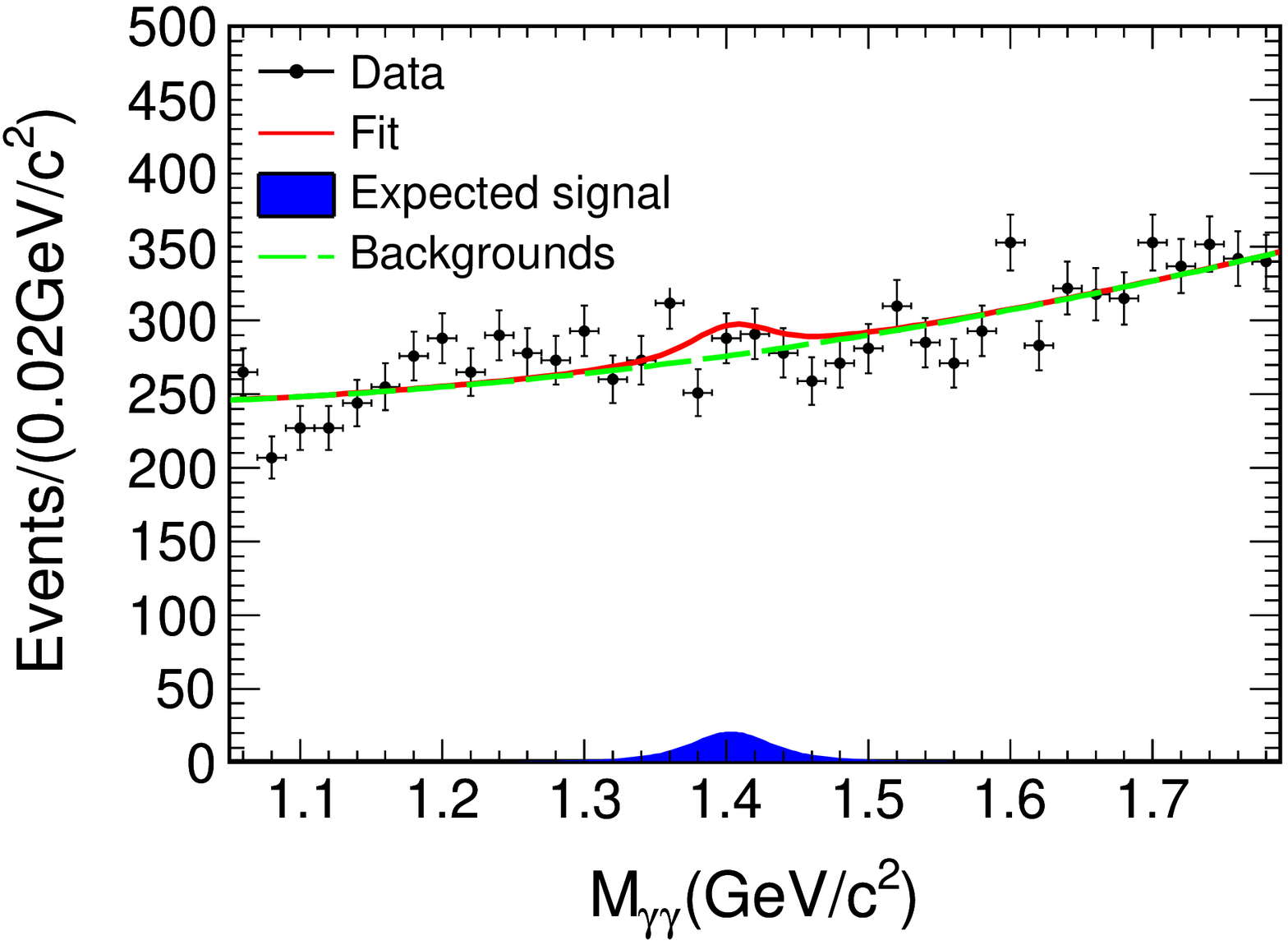}
  \put(80,60){{\bf(a)  }}
  \end{overpic}
  \begin{overpic}[width=0.4\textwidth]{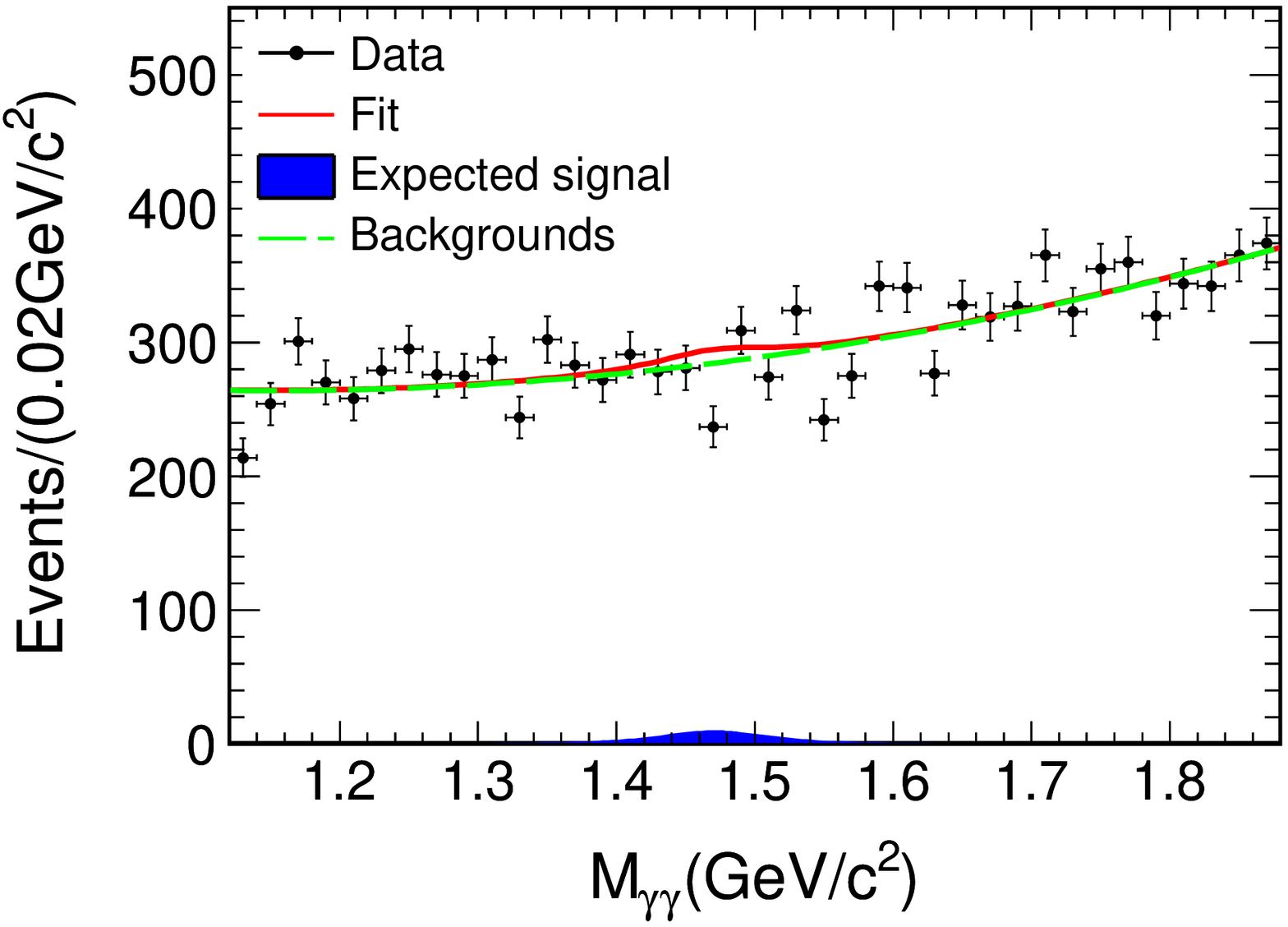}
  \put(80,60){{\bf(b)  }}
  \end{overpic}
  }

  \mbox{
  \begin{overpic}[width=0.4\textwidth]{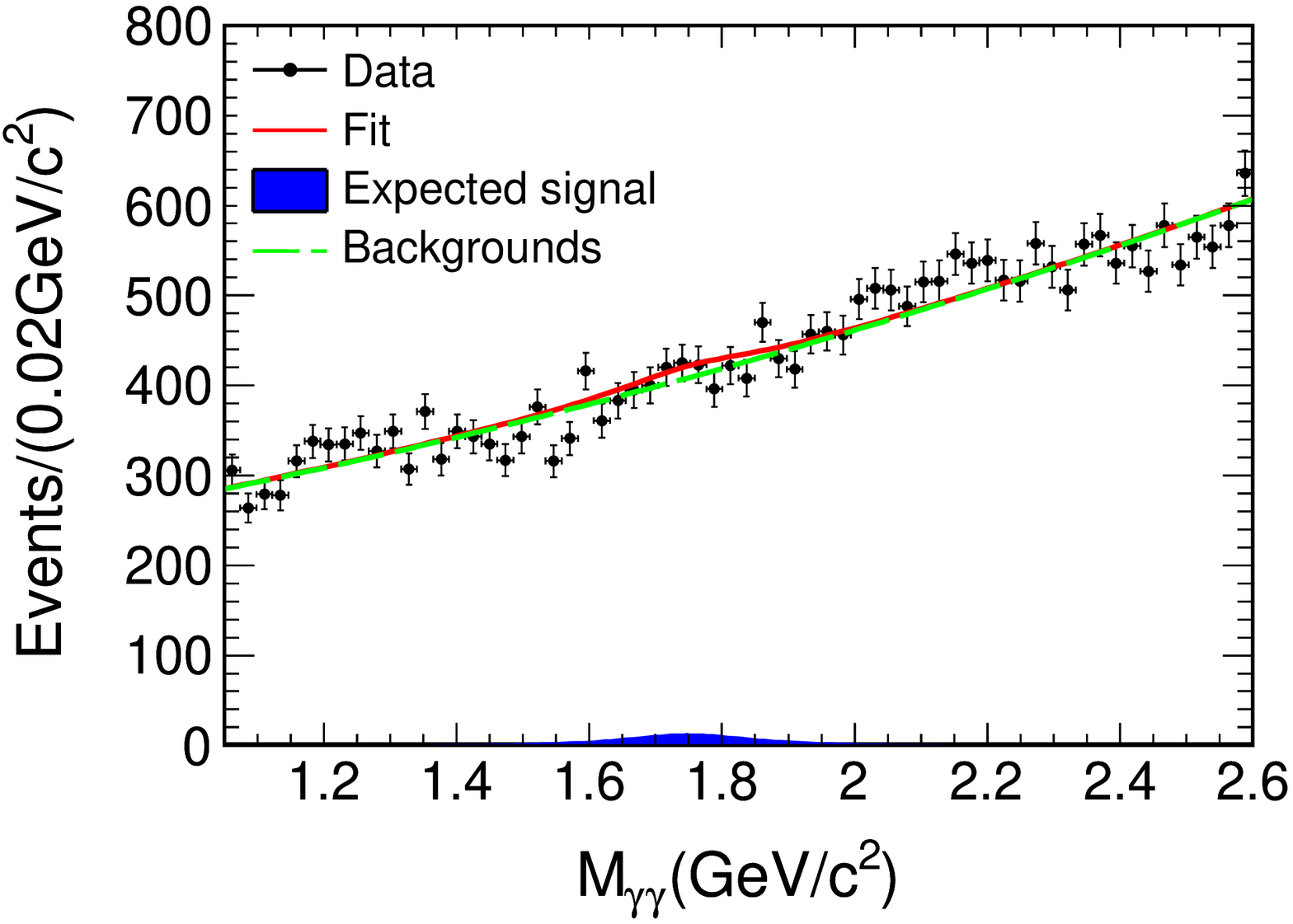}
  \put(80,60){{\bf(c)  }}
  \end{overpic}
  \begin{overpic}[width=0.4\textwidth]{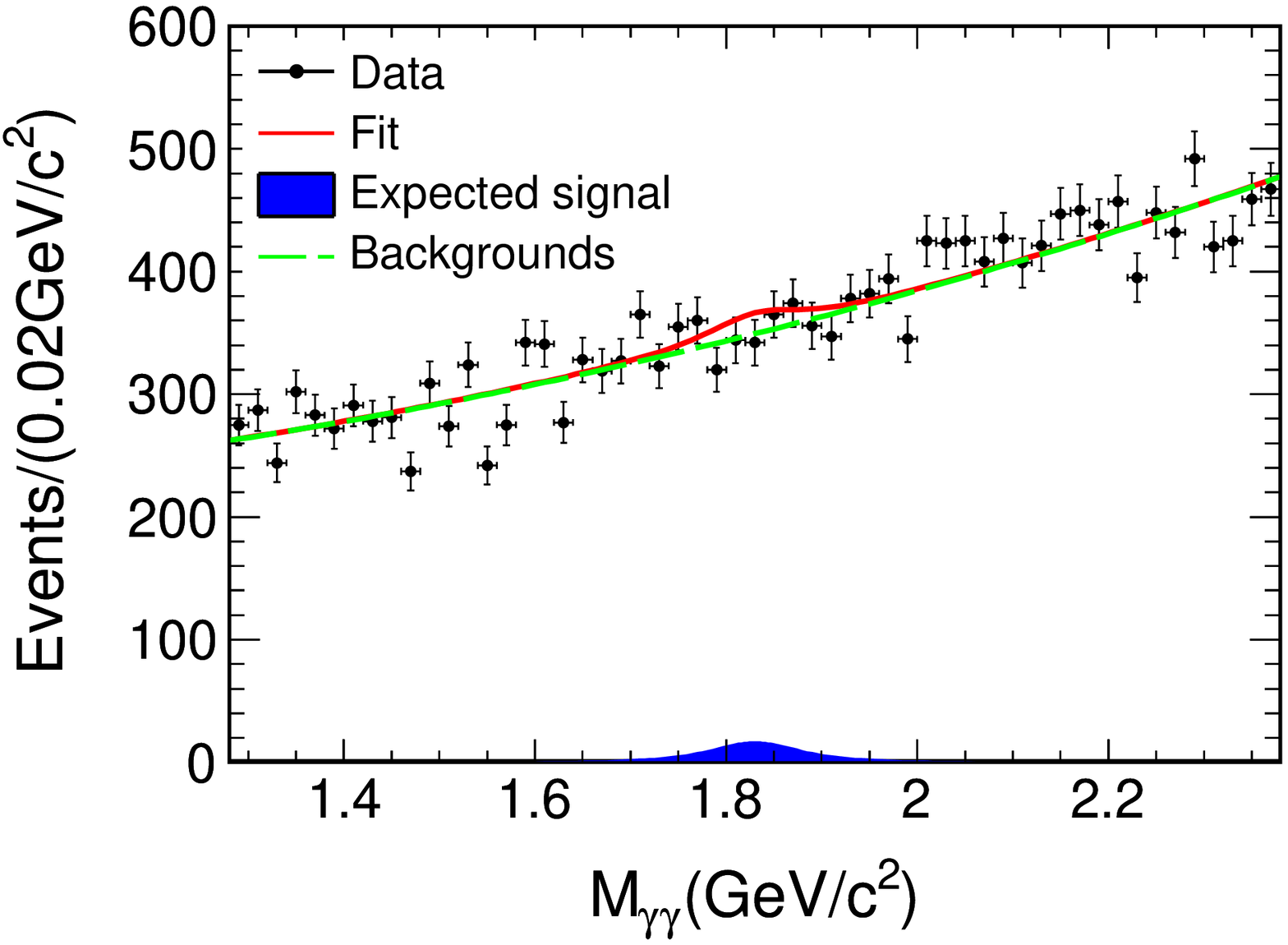}
  \put(80,60){{\bf(d)  }}
  \end{overpic}
  }
  \caption{\label{fituplimit}(color online) Fit results for the
    $\gamma\gamma$ invariant mass distributions for (a)
    $J/\psi\to\gamma\eta(1405)\to3\gamma$, (b)
    $J/\psi\to\gamma\eta(1475)\to3\gamma$, (c)
    $J/\psi\to\gamma\eta(1760)\to3\gamma$ and (d) $J/\psi\to\gamma
    X(1835)\to3\gamma$. The dots with error bars are data, the red
    solid curves show the result of the fit, the blue shaded histograms are
    the expected signals, where the signal normalization corresponds to
    the 90$\%$ confidence level upper limit, and the green long-dashed curves
    show the background.}
\end{figure*}

The signal yields of $J/\psi\to\gamma(\pi^{0},~\eta,~\eta')\to3\gamma$
are obtained from unbinned maximum likelihood fits to the
$\gamma\gamma$ invariant mass spectra.  In the fits, the signal shapes
are modeled with the sum of a CB function and a Gaussian function.  The
non-$J/\psi$ backgrounds are estimated with the events in the $J/\psi$
sideband region, assuming the $M_{\gamma\gamma\gamma}$ distribution
to be flat in the vicinity of the $J/\psi$.  Their yields and shapes are fixed in the fit.  The
non-peaking $J/\psi$ background is parameterized with a second-order Chebychev
polynomial function.  The fit results are shown in
Fig.~\ref{fitMCpi0}. The signal yields from the fit and the MC determined
detection efficiencies are summarized in Table~\ref{fitresult}, where
the MC simulation is performed using an angular distribution of
$1+\cos^{2}\theta_{\gamma}$ for the radiative photon in the $J/\psi$
rest frame.

No obvious signals for the pseudoscalar mesons $\eta(1405)$, $\eta(1475)$, $\eta(1760)$ or
$X(1835)$ are observed in the M$_{\gamma\gamma}$ distributions.
Upper limits on the signal yields are obtained by fits to the
$M_{\gamma\gamma}$ distributions in the vicinity of the corresponding
signal region, as shown in Fig.~\ref{fituplimit}. In the fits, the line
shapes of the $\eta(1405)$, $\eta(1475)$, $\eta(1760)$ and $X(1835)$
signals
are parameterized by Breit Wigner (BW) functions convolved with
Gaussian functions to account for the mass resolution, where the mass
and width of BW functions are fixed to
the world average values taken from the PDG~\cite{PDG} and the mass
resolutions are obtained from MC simulation. The background shapes are
described by second-order Chebychev polynomial functions.  We derive
the upper limits from these fits using a Bayesian approach with a flat
prior as input.  The
distribution of minimized likelihood values for a series of input
signal event yields is taken as the probability density function (PDF)
for the expected number of events.  The number of events at 90$\%$ of
the integral of the PDF from 0 to the given number of events is
defined as the upper limit at the 90$\%$ confidence level (C.L.).  To
take into account the systematic uncertainties related to the fits,
alternative fits with different fit ranges and background shapes are
also performed, and the maximum upper limit among these cases is
selected.

\begin{table*}[htbp]
  \caption{\label{fitresult}Numbers used in the calculations of the product branching fractions and the upper limits, including the numbers of events, $N_\text{obs}$($N_\text{UL}$), the detection efficiency, $\varepsilon$, and the product branching fractions, $B$. The world average values (PDG) are shown for comparison.}
\begin{ruledtabular}
 \begin{tabular}{lcccccccc}
    Decay mode     &~~$N_\text{obs}$($N_\text{UL})$       &~~$\varepsilon$(\%)    &~~$B$     &~~PDG\\ \hline
$J/\psi\to\gamma\pi^{0}\to3\gamma$ \Tstrut &~~$1635\pm54$ &~~$29.03\pm0.08$&~~$(3.57\pm0.12\pm0.16)\times10^{-5}$ &~~$(3.45^{~+~0.33}_{~-~0.30}\phantom{.})\times10^{-5}$\\
$J/\psi\to\gamma\eta\to3\gamma$    &~~$18551\pm158$    &~~$27.18\pm0.07$    &~~$(4.42\pm0.04\pm0.18)\times10^{-4}$ &~~$(4.35\pm0.14)\times10^{-4}$\\
$J/\psi\to\gamma\eta'\to3\gamma$   &~~$5057\pm94$      &~~$26.00\pm0.08$    &~~$(1.26\pm0.02\pm0.05)\times10^{-4}$ &~~$(1.14\pm0.05)\times10^{-4}$\\ 
$J/\psi\to\gamma\eta(1405)\to3\gamma$ &~~$<103$ &~~$25.37\pm0.09$  &~~$<2.63\times10^{-6}$ &~~ --\\
$J/\psi\to\gamma\eta(1475)\to3\gamma$ &~~$<~73$ &~~$25.41\pm0.11$  &~~$<1.86\times10^{-6}$ &~~ --\\
$J/\psi\to\gamma\eta(1760)\to3\gamma$ &~~$<191$ &~~$25.73\pm0.12$  &~~$<4.80\times10^{-6}$ &~~ --\\
$J/\psi\to\gamma X(1835)\to3\gamma$   &~~$<143$ &~~$25.99\pm0.11$  &~~$<3.56\times10^{-6}$ &~~ --\\
\end{tabular}
\end{ruledtabular}
\end{table*}

\section{SYSTEMATIC UNCERTAINTIES}

\begin{table*}[htbp]
  \caption{\label{Total systematic error}
    Sources of relative systematic uncertainties and their contributions to the product branching fractions and upper limits (in $\%$).}

\begin{ruledtabular}
 \begin{tabular}{lcccccccc}
      Source   &~~ $\pi^{0}$ &~~ $\eta$ &~~ $\eta'$  &~~ $\eta(1405)$ &~~ $\eta(1475)$ &~~ $\eta(1760)$ &~~ X(1835) \\ \hline
      MDC tracking      \Tstrut              &~~ 2.0 &~~2.0 &~~2.0  &~~ 2.0 &~~2.0 &~~2.0 &~~2.0\\
      Photon identification                  &~~ 3.0 &~~3.0 &~~3.0  &~~ 3.0 &~~3.0 &~~3.0 &~~3.0\\
      4C kinematic fit                       &~~ 0.4 &~~0.4 &~~0.4  &~~ 0.4 &~~0.6 &~~0.4 &~~0.5\\
      $J/\psi$ mass window                   &~~ 0.2 &~~0.2 &~~0.2  &~~ 0.2 &~~0.2 &~~0.2 &~~0.2\\
      Fit range                              &~~ 1.5 &~~0.6 &~~0.8  &~~ --  &~~ -- &~~ -- &~~ --\\
      Background shape                       &~~ 1.3 &~~1.0 &~~0.8  &~~ --  &~~ -- &~~ -- &~~ --\\
      Sideband region                        &~~ 0.9 &~~0.4 &~~0.6  &~~ --  &~~ -- &~~ -- &~~ --\\
      MC statistics                          &~~ 0.3 &~~0.3 &~~0.3  &~~ 0.4 &~~0.4 &~~0.5 &~~0.4\\
      $\psi(3686)\to\pi^{+}\pi^{-}J/\psi$    &~~ 0.9 &~~0.9 &~~0.9  &~~ 0.9 &~~0.9 &~~0.9 &~~0.9\\
      Number of $\psi(3686)$ events          &~~ 0.6 &~~0.6 &~~0.6  &~~ 0.6 &~~0.6 &~~0.6 &~~0.6\\ \hline
     Total    \Tstrut                        &~~ 4.4 &~~4.0 &~~4.0  &~~ 3.8 &~~3.9 &~~3.8 &~~3.8\\   
\end{tabular}
\end{ruledtabular}
\end{table*}

Systematic uncertainties in the branching fraction measurements mainly
originate from efficiency differences between data and MC simulation in
the MDC tracking, the photon detection, the kinematic fitting
efficiency and the $J/\psi$ mass window requirement.
Additional uncertainties associated with the fit range, the background shape,
the sideband regions, the MC statistics, the branching fraction of
$\psi(3686)\to\pi^{+}\pi^{-}J/\psi$, and the total number of $\psi(3686)$ events
are also considered.

The tracking efficiency of charged pions has been investigated using
control samples of $J/\psi\to p\bar
p\pi^{+}\pi^{-}$~\cite{MDC}.  The difference in tracking
efficiency between data and MC simulation is found to be 1$\%$ per track,
which is taken as the uncertainty from the tracking efficiency.

The photon detection efficiency is studied with a clean sample of
$J/\psi\to\rho^{0}\pi^{0}$~\cite{Photon}. The result shows that
the difference of detection efficiency between data and MC
simulation is 1\% per photon.

The systematic uncertainties associated with the 4C kinematic fit are
studied with the track helix parameter correction method, as described
in Ref.~\cite{4cError}. In this analysis, we take the efficiencies
with correction as the nominal values, and the differences with
respect to those without corrections are taken as the systematic
uncertainties associated with the 4C kinematic fit.

Due to the difference in the mass resolution between data and MC, the
uncertainty related to the $J/\psi$ mass window requirement is
investigated by smearing the MC simulation in accordance with the
signal shape of data. The change of the detection efficiency is
assigned as the systematic uncertainty for the $J/\psi$ mass window
requirement.

To study the uncertainty from the fit range, the fit is repeated
with different fit ranges, and the resultant largest differences in
the signal yields are taken as the systematic uncertainties.

To estimate the uncertainty associated with the background shape,
alternative fits with first-order or third-order Chebychev polynomial
functions for the background are performed, and the maximum differences
in signal yields with respect to the nominal values are taken as the
systematic uncertainties.

The uncertainties from the $J/\psi$ sideband region is
estimated by using alternative sideband regions.
The maximum differences
  in signal yields  are taken as the uncertainties.

The uncertainty from the decay branching fractions of
$\psi(3686)\to\pi^{+}\pi^{-}
J/\psi$ is taken from the PDG~\cite{PDG}, and the systematic
uncertainty due to the number of $\psi(3686)$ events is determined
to be 0.7$\%$ according to Ref.~\cite{NumberOfpsip0912}.

Table~\ref{Total systematic error} summarizes the systematic
uncertainties from all sources for each decay. The systematic
uncertainties associated with the statistics of MC samples are also
included. The total systematic uncertainty is obtained by adding all
individual uncertainties in quadrature, assuming all sources to be
independent.

 \section{RESULTS}

 The product branching fraction of $J/\psi\to\gamma P \to3\gamma$ is
 calculated using
\begin {equation}
\begin{gathered}
 \label{eqbrpi0}
  B(J/\psi\to\gamma P\to3\gamma)= \\
  \frac{N_\text{obs}-N_\text{bkg}}{N_{\psi(3686)}\cdot B(\psi(3686)\to\pi^{+}\pi^{-}J/\psi)\cdot\varepsilon},
\end{gathered}
\end {equation}
where $P$ represents the pseudoscalar meson, $N_\text{obs}$ is the
number of observed signal events determined from the fit to the
$\gamma\gamma$ mass spectra, $N_\text{bkg}$ is the number of peaking
background events, $N_{\psi(3686)}$ is the total number of
$\psi(3686)$ events~\cite{NumberOfpsip0912}, $\varepsilon$ is the
MC-determined detection efficiency and
$B(\psi(3686)\to\pi^{+}\pi^{-}J/\psi$) is the branching fraction of
$\psi(3686)\to\pi^{+}\pi^{-}J/\psi$~\cite{PDG}.

The product branching fractions of
$J/\psi\rightarrow\gamma(\pi^0,~\eta,~\eta^\prime)\rightarrow3\gamma$,
are then determined to be $(3.57\pm0.12\pm0.16)\times10^{-5}$,
$(4.42\pm0.04\pm0.18)\times10^{-4}$ and
$(1.26\pm0.02\pm0.05)\times10^{-4}$, respectively, as summarized in
Table~\ref{fitresult}.  For comparison we also calculate the product
branching fractions using the world average values of
$B(J/\psi\rightarrow\gamma P)$ and $B(P\rightarrow\gamma\gamma)$ from
the PDG~\cite{PDG}, and the our measured branching fractions and the PDG
branching fractions are summarized in Table~\ref{fitresult}.  The
first two branching fractions are in good agreement with the world
average values, which are dominated by the results from
BESII~\cite{BESII} and CLEO~\cite{CLEO},
while the third one is slightly higher than the world average value,
but consistent within two standard deviations.

To estimate the upper limits on product decay branching fractions for
un-observed pseudoscalar mesons, the systematic uncertainties are
taken into consideration by convolving the PDF of likelihood values in
each decay with a Gaussian function $G(\mu, \sigma) = G(0,
N \sigma_\text{sys})$, where $N$ is the signal yield and $\sigma_\text{sys}$ is
the corresponding relative systematic uncertainty listed in Table~\ref{Total
  systematic error}. The upper limits on the number of events and the
branching fractions for $J/\psi$ $\to$ $\gamma$
[$\eta(1405)$,~$\eta(1475)$,~$\eta(1760)$,~X(1835)]$\to3\gamma$ at the
90\%~C.L. are listed in Table~\ref{fitresult}.  Using the branching
fractions of $J/\psi\to\gamma\eta(1440)\to\gamma K
\bar{K}\pi$~\cite{1405dominant},
$J/\psi\to\gamma\eta(1760)\to\gamma\omega\omega$~\cite{1760N18} and
$J/\psi\to\gamma
X(1835)\to\gamma\pi^{+}\pi^{-}\eta'$~\cite{1835dominant} and their
uncertainties, the upper limits at the 90\%~C.L. for the ratios of
$\frac{B(\eta(1405)\to \gamma\gamma)}{B(\eta(1440)\to K \bar{K}\pi)}$,
$\frac{B(\eta(1475)\to \gamma\gamma)}{B(\eta(1440)\to K \bar{K}\pi)}$,
$\frac{B(\eta(1760)\to \gamma\gamma)}{B(\eta(1760)\to \omega\omega)}$
and $\frac{B(X(1835)\to \gamma\gamma)}{B(X(1835)\to
  \pi^{+}\pi^{-}\eta')}$ are determined to be $1.78\times10^{-3}$,
$1.27\times10^{-3}$, $2.48\times10^{-3}$ and $9.80\times10^{-3}$,
respectively, and are reported for the first time in $J/\psi$ decays.

\begin{acknowledgments}
The BESIII collaboration thanks the staff of BEPCII and the IHEP computing center for their strong support. This work is supported in part by National Key Basic Research Program of China under Contract No. 2015CB856700; National Natural Science Foundation of China (NSFC) under Contracts Nos. 11565006, 11235011, 11335008, 11425524, 11625523, 11635010; the Chinese Academy of Sciences (CAS) Large-Scale Scientific Facility Program; the CAS Center for Excellence in Particle Physics (CCEPP); Joint Large-Scale Scientific Facility Funds of the NSFC and CAS under Contracts Nos. U1332201, U1232107, U1532257, U1532258; CAS under Contracts Nos. KJCX2-YW-N29, KJCX2-YW-N45, QYZDJ-SSW-SLH003; 100 Talents Program of CAS; National 1000 Talents Program of China; INPAC and Shanghai Key Laboratory for Particle Physics and Cosmology; German Research Foundation DFG under Contracts Nos. Collaborative Research Center CRC 1044, FOR 2359; Istituto Nazionale di Fisica Nucleare, Italy; Koninklijke Nederlandse Akademie van Wetenschappen (KNAW) under Contract No. 530-4CDP03; Ministry of Development of Turkey under Contract No. DPT2006K-120470; National Science and Technology fund; The Swedish Research Council; U. S. Department of Energy under Contracts Nos. DE-FG02-05ER41374, DE-SC-0010118, DE-SC-0010504, DE-SC-0012069; University of Groningen (RuG) and the Helmholtzzentrum fuer Schwerionenforschung GmbH (GSI), Darmstadt; WCU Program of National Research Foundation of Korea under Contract No. R32-2008-000-10155-0.
\end{acknowledgments}

\end{document}